\documentclass[aps,prd,reprint,a4paper,showpacs,nofootinbib,
superscriptaddress,floatfix]{revtex4-2}
\usepackage{cmap}
\usepackage[T1]{fontenc}
\usepackage[utf8]{inputenc} 
\usepackage{lmodern}

\usepackage{graphicx}
\usepackage{dcolumn}
\usepackage{bm}
\usepackage{color}
\usepackage{enumitem}
\usepackage{amsmath}

\usepackage{amssymb}

\newcommand{\beq}{\begin{equation}}
\newcommand{\eeq}{\end{equation}}
\newcommand{\bea}{\begin{eqnarray}}
\newcommand{\eea}{\end{eqnarray}}

\newcommand{\dd}{\mathrm{d}}
\newcommand{\ee}{\mathrm{e}}

\usepackage{bbm}
\usepackage{amsfonts}
\usepackage{mathrsfs}
\usepackage{latexsym}
\usepackage{epsfig}
\usepackage{epstopdf}
\usepackage{epstopdf}
\usepackage{graphicx}
\usepackage{amssymb}
\usepackage{amsmath}
\usepackage{dcolumn}
\usepackage{bm}
\usepackage{color}
\usepackage{comment}
\usepackage{xcolor}
\usepackage{dsfont}

\usepackage{amsmath}
\usepackage{hyperref}
\hypersetup{
  colorlinks=true,        
  linkcolor=blue,         
  citecolor=cyan,         
}

\begin{document}

\title{\textbf{Particle Dynamics and Thermodynamics of a Charged-Like Hairy Black Hole in Extended Gravitational Decoupling}}

\author{M. Zeeshan Gul}
\email{mzeeshangul.math@gmail.com} 

\affiliation{ Tongji University, Shanghai 201804, China}

\affiliation{Department of Mathematics and Statistics, The
University of Lahore, 1-KM Defence Road Lahore-54000, Pakistan}

\affiliation{Research Center of Astrophysics and Cosmology, Khazar
University, Baku, AZ1096, 41 Mehseti Street, Azerbaijan}

\affiliation{Jadara Research Center, Jadara University, Irbid 21110, Jordan}

\author{Ahmad Al-Badawi}
\email{ahmadbadawi@ahu.edu.jo}
\affiliation{Department of Physics, Al-Hussein Bin Talal University, P. O. Box: 20, 71111,
Ma'an, Jordan}

\author{J. Andrade}
\email{julio.andrade@espoch.edu.ec}
\affiliation{Facultad de Ciencias, Escuela Superior Polit\'ecnica de Chimborazo (ESPOCH), 060155 Riobamba, Ecuador}

\author{Sadia Zahid}
\email{sadiazahid.math@gmail.com} \affiliation{Department of
Mathematics and Statistics, The University of Lahore, 1-KM Defence
Road Lahore-54000, Pakistan}

\begin{abstract}
This work focuses on the analysis of particle dynamics and the thermodynamic properties associated with a particular branch of hairy black hole solutions. These solutions are obtained by using extended geometric deformation gravitational decoupling on the seed solution of the Schwarzschild solution. This resulting geometry satisfies the dominant energy condition and the condition $Q^{2}=2\chi\ell M$. We analyze the motion of massive particles and photons in this geometry and the thermodynamics of this solution. In particular, for massless particles, we study the effective potential and determine the radii of the photon sphere and the shadow of the black hole. For massive particles, we calculate the specific energy and angular momentum of circular orbits, the ISCO, the radiative efficiency of an accretion disk, and representative particle trajectories. Finally, we investigate the thermodynamic properties of this  hairy black hole geometry, including the Hawking temperature, the Bekenstein-Hawking entropy, and the heat capacity in the fixed-$(Q,\ell)$ ensemble.\end{abstract}

\maketitle

\section{Introduction}\label{intro}

Black holes (BHs) rank among the most intriguing physical systems in the
Universe, as they provide a setting for exploring extreme gravitational
regimes and serve as natural laboratories for investigating possible
connections between gravity and quantum phenomena. The conceptual
origins of BHs can be traced back to the eighteenth-century
notion of ``dark stars,'' independently discussed by John Michell and
Pierre-Simon Laplace within Newtonian gravity
\cite{michell1784vii, marquis1824exposition, montgomery2009michell}.
These Newtonian constructions, however, were not BHs in the
modern relativistic sense, since their description involved neither
spacetime curvature nor an event horizon. The modern relativistic
description emerged from Einstein's field equations (EFEs) and the
first exact, static, and spherically symmetric vacuum solution obtained
by Schwarzschild in 1916 \cite{schwarzschild1916gravitationsfeld}. These theoretical predictions have acquired strong observational support over the past few decades. First, spectroscopic observations of the Cygnus X-1 binary system revealed an invisible massive companion, establishing it as one of the first and most compelling candidates for a stellar-mass BH \cite{webster1972cygnus, bolton1972identification}. Subsequently, the detection of gravitational waves from mergers of compact binary systems and the reconstruction of images of M87* and Sgr~A* have opened up new possibilities for testing the geometry surrounding astrophysical BHs \cite{abbott2016ligo, event2019first, event2022first}. In this context, photon spheres, BH shadows, innermost stable circular orbits (ISCOs), and accretion efficiencies are useful theoretical diagnostic tools for detecting deviations from the geometries of BH solutions such as Schwarzschild, Kerr, and others. Their study can reveal how additional fields or effective gravitational sectors modify light propagation and the motion of matter near the event horizon. 

Initially, these objects were thought to follow a standard paradigm of ``hairless'' BHs; that is, stationary BHs in Einstein-Maxwell theory are characterized by their mass, angular momentum, and electric charge \cite{mazur1982proof, heusler1998stationary, chrusciel2012stationary}. However, this assumption is based on specific assumptions about matter content, asymptotic behavior, and spacetime regularity. Thus, several extensions of general relativity, as well as Einstein's gravity coupled to additional fields, admit solutions for BHs with supplementary charges or nontrivial matter distributions outside the horizon \cite{bizon1990colored, kanti1996dilatonic, herdeiro2014kerr, sotiriou2014black, herdeiro2015asymptotically}. These configurations are commonly known as ``hairy'' BHs and provide a phenomenological framework for investigating possible deviations from general relativity's vacuum. In this context, gravitational decoupling (GD) provides a systematic procedure for generating such configurations \cite{ovalle2018black}. In this theoretical framework, the total energy-momentum tensor that appears in the field equations is divided into simpler sectors, allowing a known seed geometry to be deformed by an additional source. This geometric deformation can affect only the radial metric component, in which case it is called minimal geometric deformation (MGD) \cite{ovalle2017decoupling}, or it can be implemented through extended geometric deformation (EGD) \cite{ovalle2019decoupling}, which deforms both the radial and temporal metric components. The latter is particularly suitable for BH geometries, as it allows us to obtain solutions for hairy BHs with well-defined event horizons.

Indeed, the use of GD through EGD has made it possible to obtain a family of hairy BHs by imposing the existence of a well-defined event horizon together with strong or dominant energy conditions outside the horizon \cite{ovalle2021hairy}. These solutions are characterized by having gravitational hair encoded by the decoupling parameter $\chi$ and a length scale $\ell$, associated with the primary hair. In the case of solutions that satisfy the strong energy condition (SEC), the hair is characterized by these parameters. By contrast, the general family satisfying the dominant energy condition (DEC) incorporates an additional effective charge parameter $Q$, which gives the resulting geometry type
Reissner--Nordstr\"om. The geodesic structure of these hairy BHs, including their effective potentials, photon spheres, impact parameters, and innermost stable circular orbits, has been previously investigated in \cite{ramos2021geodesic}. Their near-horizon thermodynamics and thermal stability have also been examined in \cite{cavalcanti2022near, mahapatra2023rotating}. These results necessitate the precise identification of the solution branch and observables considered in any subsequent analysis.

In this work, we focus on a specific charged-like DEC branch characterized by the exponentially decaying deformation term $-\chi M e^{-r/M}/r$ and the constraint
\begin{equation}\label{Q-condition}
Q^{2}=2\chi\ell M.
\end{equation}
Here, $\chi$ controls the geometric deformation and $\ell$ is the associated hair scale. Rather than deriving the entire family of gravitationally decoupled BHs again, our aim is to provide a unified analysis of the optical, orbital, and thermal properties of this particular branch. We first determine the physically admissible parameter region by requiring the existence of an outer event horizon and the fulfillment of the DEC. This restriction is essential because not every formal choice of $(\chi,\ell)$ represents a BH supported by the matter sector used in its construction. Thus, in this manuscript, we investigate null and timelike geodesics. In the null sector, we analyze the effective potential, effective radial force, photon sphere radius, and corresponding shadow radius. In the timelike sector, we study circular motion, specific energy and angular momentum, the ISCO radius, and radiative efficiency. Numerical trajectories are also obtained to illustrate the effects of deformation parameters on the orbital motion. Finally, we examine the Hawking temperature and Bekenstein--Hawking entropy and perform a heat-capacity analysis.

The article is organized as follows. Section \eqref{section II} summarizes the extended gravitational decoupling formalism. Section \eqref{section III} introduces the geometry of the hair-structured BH and establishes its domain of physical parameters. We then study null and timelike geodesics, photon spheres, shadows, circular orbits, and particle trajectories in Section \eqref{section IV}. Section \eqref{section V} analyzes the thermodynamic properties of the solution. Finally, Section \eqref{section VI} summarizes our main results and conclusions.

\section{Gravitational decoupling by EGD}\label{section II}

In this section, we will briefly review the GD using EGD for those gravitational systems with spherical symmetry and static described (for more details see reference \cite{ovalle2019decoupling}). Let us first consider the EFEs (we use geometrized units, \(c=G_N=1\), so that \(k^2=8\pi\), where $G_{\rm N}$ is Newton's constant)
\begin{eqnarray}\label{1}
G_{\beta\alpha}\equiv
R_{\beta\alpha}-\frac{1}{2}R\,g_{\beta\alpha}=k^2\mathcal{T}_{\beta\alpha},
\end{eqnarray}
with a total energy-momentum tensor containing two contributions,
\begin{eqnarray}\label{2}
\mathcal{T}_{\beta\alpha}=T_{\beta\alpha}+\Upsilon_{\beta\alpha},
\end{eqnarray}
where $T_{\beta\alpha}$ is usually associated with some known solution of general relativity (usually such a solution is an isotropic fluid), whereas $\Upsilon_{\beta\alpha}$ is an extra source that may contain new fields or a new gravitational sector. Since the Einstein tensor $G_{\beta\alpha}$ satisfies the Bianchi identity, the total source must be covariantly conserved,
\begin{eqnarray}\label{3}
\nabla^{\beta} \mathcal{T}_{\beta\alpha}=0.
\end{eqnarray}
For spherically symmetric and static systems, the metric $g_{\beta\alpha}$ can be written as
\begin{eqnarray}\label{4}
\dd s^2=\ee^{\varsigma(r)}\dd t^2-\ee^{\mu(r)}\dd r^2-r^2\dd\Omega^2,
\end{eqnarray}
where $\varsigma=\varsigma(r)$ and $\mu=\mu(r)$ are functions of the areal
radius $r$ only, and $\dd\Omega^2=\dd\theta^2+\sin^2\theta\,\dd\phi^2$. The EFEs (\ref{1}) then read
\begin{eqnarray}
\label{5} k^2\left(T^0_{\ 0}+\Upsilon^0_{\ 0}\right)
&=&\frac{1}{r^2}-\ee^{-\mu}\left(\frac{1}{r^2}-\frac{\mu'}{r}\right),\\
\label{6} k^2\left(T^1_{\ 1}+\Upsilon^1_{\ 1}\right)
&=&\frac{1}{r^2}-\ee^{-\mu}\left(\frac{1}{r^2}+\frac{\varsigma'}{r}\right),\\
\label{7} k^2\left(T^2_{\ 2}+\Upsilon^2_{\ 2}\right)
&=&-\frac{\ee^{-\mu}}{4}\bigg(2\varsigma''+\varsigma'^2-\mu'\varsigma'\nonumber\\ 
&& +2\frac{\varsigma'-\mu'}{r}\bigg),
\end{eqnarray}
where $f'\equiv \partial_r f$ and $\mathcal{T}^3_{\
3}=\mathcal{T}^2_{\ 2}$ due to the spherical symmetry. Thus, we can identify in Eqs.~ \eqref{5}--\eqref{7} an effective density
\begin{eqnarray}\label{8}
\varrho=T^0_{\ 0}+\Upsilon^0_{\ 0},
\end{eqnarray}
an effective radial pressure
\begin{eqnarray}\label{9}
\tilde p_r=-T^1_{\ 1}-\Upsilon^1_{\ 1},
\end{eqnarray}
and an effective tangential pressure
\begin{eqnarray}\label{10}
\tilde p_t=-T^2_{\ 2}-\Upsilon^2_{\ 2}.
\end{eqnarray}
Moreover, the anisotropy
\begin{eqnarray}\label{11}
\Delta\equiv \tilde p_t-\tilde p_r,
\end{eqnarray}
usually does not vanish and the system of Eqs.~\eqref{5}--\eqref{7} may be
treated as an anisotropic fluid \cite{herrera1997local, mak2003anisotropic}.

We next consider a solution to the Eqs.~\eqref{1} for the seed source
$T_{\beta\alpha}$ alone (that is, without extra source $\Upsilon_{\beta\alpha}=0$), which we write as
\begin{eqnarray}\label{12}
\dd s^2=\ee^{\eta(r)}\dd t^2-\ee^{\upsilon(r)}\dd r^2-r^2\dd\Omega^2,
\end{eqnarray}
where
\begin{eqnarray}\label{13}
\ee^{-\upsilon(r)}\equiv 1-\frac{k^2}{r}\int_0^r x^2 T^0_{\ 0}(x)\,\dd
x =1-\frac{2m(r)}{r},
\end{eqnarray}
where $m=m(r)$ is the Misner-Sharp mass function . The addition of the source $\Upsilon_{\beta\alpha}$ can then be accounted for by the EGD of the seed metric \eqref{12}, namely
\begin{eqnarray}
\label{14}
\eta \to \varsigma&=&\eta+\chi g_{1},\\
\label{15} \ee^{-\upsilon} \to \ee^{-\mu}&=&\ee^{-\upsilon}+\chi g_{2},
\end{eqnarray}
where $g_{2}$ and $g_{1}$ are known as geometric deformations for the radial and temporal metric components, and the parameter $\chi$ controls the intensity of such deformations on seed source. Now, using the Eqs.~\eqref{14} and \eqref{15}, the Einstein equations \eqref{5}--\eqref{7} are separated in two sets, one related to standard EFE for the source of $T_{\beta\alpha}$, that is
\begin{eqnarray}
\label{16}
k^2T^0_{\ 0}&=&\frac{1}{r^2}-\ee^{-\upsilon}\left(\frac{1}{r^2}
-\frac{\upsilon'}{r}\right),\\
\label{17}
k^2T^1_{\ 1}&=&\frac{1}{r^2}-\ee^{-\upsilon}\left(\frac{1}{r^2}
+\frac{\eta'}{r}\right),\\
\label{18} k^2T^2_{\
2}&=&-\frac{\ee^{-\upsilon}}{4}\left(2\eta''+\eta'^2-\upsilon'\eta'
+2\frac{\eta'-\upsilon'}{r}\right),
\end{eqnarray}
which is assumed to be solved by the seed metric \eqref{12}; a second
set for the source $\Upsilon_{\beta\alpha}$ 
\begin{eqnarray}
\label{19}
k^2\Upsilon^0_{\ 0}&=&-\chi\frac{g_{2}}{r^2}-\chi\frac{g_{2}'}{r},\\
\label{20} k^2\Upsilon^1_{\ 1}+\chi \mathcal{Z}_1&=&-\chi
g_{2}\left(\frac{1}{r^2}+
\frac{\varsigma'}{r}\right),\\
\nonumber k^2\Upsilon^2_{\ 2}+\chi
\mathcal{Z}_2&=&-\chi\frac{g_{2}}{4}\left(2\varsigma''+\varsigma'^2+2\frac{\varsigma'}{r}\right)
-\chi\\\label{21}&\times& \frac{g_{2}'}{4}\left(\varsigma'+\frac{2}{r}\right),
\end{eqnarray}
where
\begin{eqnarray}
\mathcal{Z}_1 &=& \frac{\ee^{-\upsilon}g_{1}'}{r},\ \text{and}\label{22}\\
\mathcal{Z}_2 &=& \frac{\ee^{-\upsilon}}{4}\left(2g_{1}''+g_{1}'^2+\frac{2g_{1}'}{r}+2\eta'g_{1}'-\upsilon'g_{1}'\right).\label{23}
\end{eqnarray}
Precisely because the EFEs (\ref{1}) can be separated into these two subsets, this theoretical method is called ``gravitational decoupling''. Also, note that the above equations clearly show that the effect of the extra source represented by the tensor $\Upsilon_{\beta\alpha}$ must vanish when the deformations vanish ($\chi=0$). 

\section{Hairy black holes}\label{section III}

Several scenarios and conditions have been studied in which it can be stated that a black hole can have hair \cite{sotiriou2012black, babichev2014dressing, cisterna2014asymptotically, antoniou2018evasion, antoniou2018black, grumiller2020spacetime, volkov1989non, kanti1998winstanley, zloshchastiev2005coexistence}. Among these possible scenarios, the possibility of filling the static void with some potentially fundamental source stands out for its simplicity, often described through a scalar field \cite{martinez2004exact, sotiriou2015black}. Ovalle et al. \cite{ovalle2021hairy} considered a more general scenario in which the Schwarzschild vacuum is surrounded by a generic static and spherically symmetric tensor-vacuum. This implies hairy black hole solutions described by the mass $M$ and a discrete set of charges generating primary hair. However, the MGD leaves the temporal component of the metric \eqref{4} exactly equal to the Schwarzschild one
\begin{eqnarray}\label{28}
\ee^\varsigma=\ee^\eta=1-\frac{2M}{r},
\end{eqnarray}
which hinders the existence of stable BHs with a well-defined event horizon. The relation \eqref{28}. The advantage of the EGD is that the time component is also modified according to equation Eq.~\eqref{14}, which produces a potentially larger number of hairy BH solutions with horizons other than $r_h=2M$. We apply the analysis from the previous section to the particular case of $T_{\beta\alpha}=0$. The seed metric \eqref{12} is given by the Schwarzschild solution with
\begin{eqnarray}\label{29}
\ee^\eta=\ee^{-\upsilon}=1-\frac{2M}{r},
\end{eqnarray}
which solves Eqs.~\eqref{16}--\eqref{18} for $T_{\beta\alpha}=0$. In order to
find hairy BHs, we then need to solve the resulting ``quasi-Einstein'' system \eqref{19}--\eqref{21}, which contain the three components of $\Upsilon_{\beta\alpha}$ and the two deformations $g_{2}$ and $g_{1}$. Furthermore, we reduce the number of unknown quantities, so that they specify a unique solution, by prescribing the two conditions discussed in the Introduction.

First of all, in order to have BH solutions with a well-defined horizon structure, we require the deformed metric \eqref{4} to satisfy 
\begin{eqnarray}\label{30}
\ee^\varsigma=\ee^{-\mu}.
\end{eqnarray}
We remark that Eq.~\eqref{30} implies the condition \eqref{31} but is not necessary for it to hold, as one could also consider cases with $\ee^\varsigma\ne\ee^{-\mu}$ for $r\ne r_h$. This condition ensures that the radius $r=r_h$ such that
\begin{eqnarray}\label{31}
\ee^{\varsigma(r_h)}=\ee^{-\mu(r_h)}=0,
\end{eqnarray}
will be both a Killing horizon ($\ee^\varsigma=0$) and a causal horizon
($\ee^{-\mu}=0$). A direct consequence of the condition \eqref{30},
following from the EFEs \eqref{5} and \eqref{6}, is the equation
of state
\begin{eqnarray}\label{32}
\tilde p_r=-\varrho.
\end{eqnarray}
Therefore, the radial pressure must be exclusively negative to ensure that the energy density is always positive. The condition \eqref{30} and the Schwarzschild solution \eqref{29} then relate the metric deformations $g_{2}$ and $g_{1}$ according to
\begin{eqnarray}\label{33}
\chi g_{2}(r)=\left(1-\frac{2M}{r}\right)\left(\ee^{\chi
g_{1}(r)}-1\right),
\end{eqnarray}
so that the line element \eqref{4} becomes
\begin{widetext}
\begin{eqnarray}\label{34}
\dd s^2=\left(1-\frac{2M}{r}\right)\ee^{\chi g_{1}(r)}\dd t^2
-\left(1-\frac{2M}{r}\right)^{-1}\ee^{-\chi g_{1}(r)}\dd
r^2-r^2\dd\Omega^2.
\end{eqnarray}
\end{widetext}
We are now left with the deformation $g_{1}$ and the three components of $\Upsilon_{\beta\alpha}$, which necessarily satisfies the three “quasi-Einstein” Eqs.~\eqref{19}--\eqref{21}. Therefore, it is possible to impose a physically consistent constraint on $g_{1}$ or a constraint on $\Upsilon_{\beta\alpha}$, like a reasonable equation of state. This is correct since the energy conditions on the energy-momentum tensor ensure that we have physically well-behaved and non-exotic gravitational sources \cite{visser1995lorentzian, curiel2017primer}. In particular, it should be noted that, in Ref. \cite{ovalle2021hairy}, the strong energy condition (SEC) and the dominant energy condition (DEC) were imposed on the source $\Upsilon_{\beta\alpha}$ only in the spacetime region accessible to an external observer, while possible violations of these conditions were allowed inside the event horizon. However, in this work, we focus exclusively on the DEC branch, as mentioned in the Introduction.

\subsection{Dominant energy condition}

We shall next consider the DEC, which requires \cite{visser1995lorentzian}
\begin{eqnarray}\label{75}
\varrho&\ge |\tilde p_r|,\\
\label{76} \varrho&\ge |\tilde p_t|.
\end{eqnarray}

We first point out that the inequality \eqref{75} is saturated as a
consequence of Eq.~\eqref{32} for a positive effective density, for which
Eq.~\eqref{76} reduces to
\begin{eqnarray}\label{77}
-\varrho\le \tilde p_t\le \varrho.
\end{eqnarray}
We can again write the condition \eqref{77} in terms of the definitions
\eqref{8} and \eqref{10} as
\begin{eqnarray}
\label{78}
\Upsilon^0_{\ 0}+\Upsilon^2_{\ 2}&\ge 0,\\
\label{79} \Upsilon^0_{\ 0}-\Upsilon^2_{\ 2}&\ge 0,
\end{eqnarray}
which yield respectively the differential inequalities
\begin{eqnarray}
\nonumber
H_1(r)&\equiv& -r(r-2M)h''-4(r-M)h'\\\label{80}&-&2h+2\ge 0,\\
\label{81} H_2(r)&\equiv& r(r-2M)h''+4Mh'-2h+2\ge 0,
\end{eqnarray}
where we used Eqs.~\eqref{19} and \eqref{21}, and 
\begin{eqnarray}\label{37}
h(r)=\ee^{\chi g_{1}(r)}.    
\end{eqnarray}

Upon solving \eqref{80} for the corresponding $h$, we obtain (see Ref. \cite{ovalle2021hairy} for details of calculations)
\begin{eqnarray}\label{84}
h(r)=1-\frac{1}{r-2M}\left(\chi\ell+\chi
M\ee^{-r/M}-\frac{Q^2}{r}\right),
\end{eqnarray}
where $\ell$ denotes a length scale associated with the gravitational hair and $Q$ is an effective charge parameter. For the charged-like branch considered in this work, these parameters satisfy Eq.~\eqref{Q-condition}, which ensures that $Q\rightarrow0$ and the Schwarzschild geometry is recovered in the hairless
limit $\chi\rightarrow0$. A second constant of integration was adjusted to meet the proper Schwarzschild limit for $\chi\to0$ (in which we remark that $Q^{2}\sim\chi$ vanishes as well). The deformation in Eq.~\eqref{84} also has to satisfy the inequality \eqref{81}, which reads
\begin{eqnarray}\label{85}
\frac{4Q^2}{r^2}\ge \frac{\chi}{M}(r+2M)\ee^{-r/M}.
\end{eqnarray}
Using \eqref{84} in the line element \eqref{34}, we obtain the metric functions
\begin{eqnarray}\label{86}
\ee^\varsigma=\ee^{-\mu}=1-\frac{2M+\chi\ell}{r}+\frac{Q^2}{r^2}-\frac{\chi
M\ee^{-r/M}}{r}.
\end{eqnarray}
The term $Q^{2}/r^{2}$ gives this geometry a Reissner--Nordstr\"om-like
character, although $Q$ represents an effective charge rather than
necessarily an electromagnetic one. Indeed, in the asymptotic region,
the metric function behaves as
\begin{eqnarray}
f(r)=1-\frac{2\mathcal{M}}{r}+\frac{Q^{2}}{r^{2}}+\mathcal{O}\!\left(\frac{e^{-r/M}}{r}\right),    
\end{eqnarray}
where
\begin{eqnarray}
\mathcal{M}=M+\frac{\chi\ell}{2},    
\end{eqnarray}
is the asymptotic mass \cite{ovalle2021hairy}.

With the information of the metric \eqref{86} in EFE \eqref{5}-\eqref{7}, we obtain the effective density given by
\begin{eqnarray}\label{87}
\varrho=\Upsilon^0_{\ 0}=-\tilde
p_r=\frac{Q^2}{k^2r^4}-\frac{\chi\ee^{-r/M}}{k^2r^2},
\end{eqnarray}
and an effective tangential pressure reads
\begin{eqnarray}\label{88}
\tilde p_t=-\Upsilon^2_{\
2}=\frac{Q^2}{k^2r^4}-\frac{\chi\ee^{-r/M}}{2k^2Mr}.
\end{eqnarray}
We can see that
\begin{eqnarray}\label{89}
\varrho-\tilde p_t=\frac{\chi\ee^{-r/M}}{2k^2Mr^2}(r-2M),
\end{eqnarray}
Equation~\eqref{89} shows that the inequality $\varrho-\widetilde{p}_{t}\geq0$ is satisfied for $r\geq2M$. However, the full DEC also requires $\varrho+\widetilde{p}_{t}\geq0$, which is equivalent to inequality~\eqref{85}.

We can also see that the physical singularity at $r=0$ remains, and the horizon radii $r_h$ are given by solutions of
\begin{eqnarray}\label{90}
\chi\ell=r_h-2M+\frac{Q^2}{r_h}-\chi M\ee^{-r_h/M}.
\end{eqnarray}

In the present work, we restrict the general DEC solution \eqref{86} to
the charged-like branch defined by Eq. \eqref{Q-condition}. Therefore, the metric function becomes
\begin{equation}\label{METRIC}
f(r)\equiv e^{\zeta}=e^{-\mu} = 1-\frac{2M+\chi\ell}{r} +\frac{2\chi\ell M}{r^{2}} -\frac{\chi M}{r}e^{-r/M}.
\end{equation}
This is the specific geometry whose geodesic and thermodynamic
properties are investigated throughout the remainder of this work.

For $\chi>0$, substituting $Q^{2}=2\chi\ell M$ into inequality~\eqref{85} and introducing the dimensionless radial coordinate $x=r/M$, we obtain
\begin{equation}
\frac{\ell}{M}\geq \frac{x^{2}(x+2)e^{-x}}{8}.
\end{equation}
The right-hand side reaches its maximum at
\begin{equation}
x_{\mathrm{max}}=\frac{1+\sqrt{17}}{2}\simeq2.5616,
\end{equation}
which yields
\begin{equation}
\frac{\ell}{M}\geq0.2888.
\end{equation}
This restriction guarantees that the DEC is satisfied throughout the exterior region considered in the following analysis.

\begin{figure}[t]
    \centering    \includegraphics[width=0.8\linewidth]{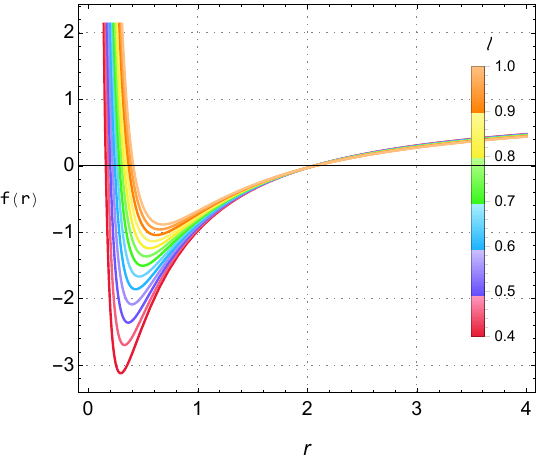}\qquad   \includegraphics[width=0.8\linewidth]{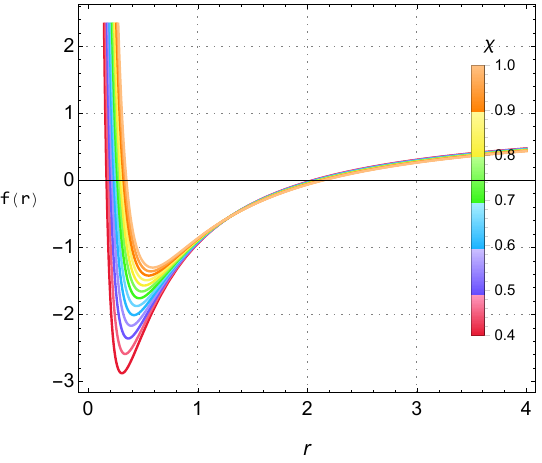}
    \caption{\footnotesize Behavior of the metric function $f(r)$ as a function of radial distance for varying EGD parameters $(\ell,\chi)$. The upper panel corresponds to fixed $\chi=0.5$, while the lower panel corresponds to fixed $\ell/M=0.5$. Here, $M=1$ is the seed mass.}
    \label{fig:Lapse}
\end{figure}

Figure \ref{fig:Lapse} shows the behavior of the metric function $f(r)$, given in Eq. (\ref{METRIC}) with $Q^2=2\chi\ell M$, as a function of the radial distance $r$ for fixed mass $M=1$ and varying EGD parameters $(\ell,\chi)$. The figure is split into two panels. In both, all curves converge toward a common path at larger radial distances ($r \ge 2$), smoothly rising to cross the horizontal axis $f(r) = 0$ near $r = 2$ and approaching positive values as $r$ increases. At smaller radii ($r < 1$), the curves diverge sharply. It is clear that the hairy BH in EGD has inner and outer horizons. The Schwarzschild horizon $r_h=2M$ is recovered only in the hairless limit $\chi=0$. 

\section{Dynamics of Test Particles}\label{section IV}

The motion of test particles provides valuable information about the physical and observational properties of BH spacetimes. Null geodesics determine photon trajectories, BH shadows, and gravitational lensing, whereas timelike geodesics describe the motion of massive particles and are closely related to accretion processes and orbital stability. In this section, we investigate both massless and massive particle motion in the spacetime described by the metric (\ref{4}).

\subsection{Metric and Geodesic Equations}

We now investigate null and timelike geodesics in the spacetime defined by Eq.~\eqref{METRIC}. Then, to derive the geodesic equations we may
utilize the Lagrangian density function \cite{ahmed2025dynamics, ahmed2025some, ahmed2025probing, al2025new, ahmed2025ads, ahmed2025photon}. Therefore, the geodesic motion in the equatorial plane ($\theta=\pi/2$) can be derived from the Lagrangian
\begin{equation}\label{p4}
\mathcal{L} = \frac{1}{2}g_{\beta\alpha}\dot{x}^{\beta}\dot{x}^{\alpha} = \frac{1}{2}\left[f(r)\dot{t}^2 - \frac{\dot{r}^2}{f(r)} - r^2\dot{\phi}^2\right],
\end{equation}
where the dot denotes differentiation with respect to an affine parameter $\lambda$. Since the metric is independent of the coordinates $t$ and $\phi$, we have two conserved quantities as
\begin{equation}\label{p5}
E = f(r)\dot{t} ,
\end{equation}
\begin{equation}\label{p6}
\mathds{L} = r^2\dot{\phi}.
\end{equation}
\subsection{Null Geodesic Equation and Effective Potential}

\begin{figure}[t]
    \centering
    \includegraphics[width=0.8\linewidth]{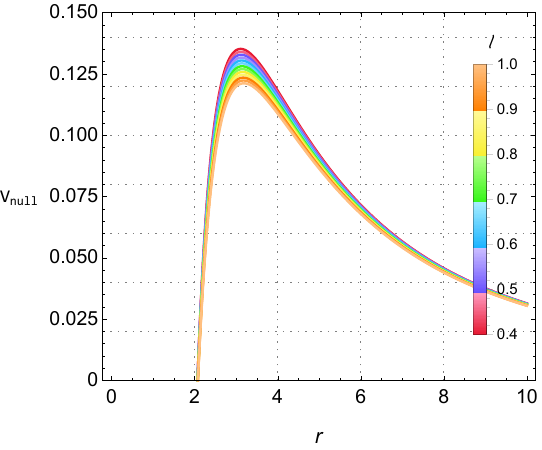}\qquad    \includegraphics[width=0.8\linewidth]{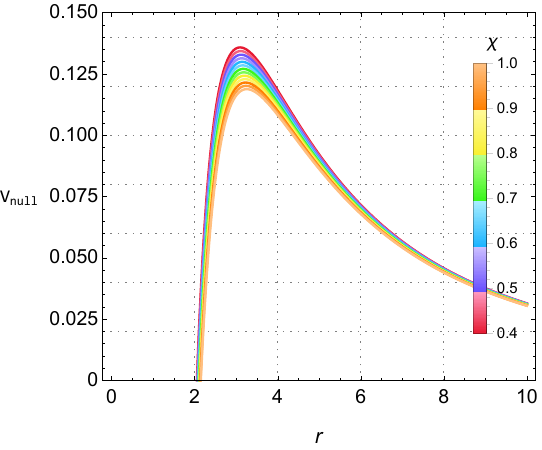}
    \caption{\footnotesize Behavior of the null effective potential
$V_{\rm null}(r)$ as a function of radial distance for varying
EGD parameters $(\ell,\chi)$. The upper panel corresponds to
fixed $\chi=0.5$, while the lower panel corresponds to fixed
$\ell/M=0.5$, with $M=1$ and $L=2$. The dashed vertical lines
mark the outer event horizons, and only the exterior region
$r\geq r_h$ is shown.}
    \label{fig:potential-null}
\end{figure}

For null geodesics, we have $ds^2=0$, which yields
\begin{equation}\label{p7}
f(r)\dot{t}^2 - \frac{\dot{r}^2}{f(r)} - r^2\dot{\phi}^2 = 0.
\end{equation}
Substituting the conserved quantities (\ref{p5}) and (\ref{p6}) into (\ref{p7}), we obtain the radial equation
\begin{equation}\label{p8}
\dot{r}^2 = E^2 - \frac{L^2}{r^2}f(r).
\end{equation}
Introducing the impact parameter
\begin{equation}\label{p9}
b_c \equiv \frac{L}{E},
\end{equation}
the radial equation can be rewritten as
\begin{equation}\label{p10}
\dot{r}^2 = E^2\left[1 - \frac{f(r)}{r^2}b_c^2\right].
\end{equation}
In general, the effective potential is defined as 
\begin{equation}\label{p11a}
V_{null}(r) =  \left(\frac{L^2}{r^2}-\xi \right)f(r) .
\end{equation}
Here $\xi=0$ for photons and $\xi=-1$ for massive particles.
Thus, for photon dynamics the effective potential is given by
\begin{eqnarray}\label{p11}
V_{null}(r) =  \frac{L^2}{r^2}f(r) &=&  \frac{L^2}{r^2}\bigg(1-\frac{2M+\chi\ell}{r}+\frac{2\chi\ell M}{r^2}\nonumber\\ 
&& - \frac{\chi M}{r}\ee^{-r/M} \bigg).
\end{eqnarray}
Fig. \ref{fig:potential-null}, shows the behavior of the effective potential for null geodesics as a function of the radial coordinate, under variations of the EGD parameters $(\ell, \chi)$. The null effective potential ($V_{\text{null}}$) exhibits a characteristic potential barrier with a peak near the unstable photon sphere, while vanishing at the outer event horizon ($r=r_h$) and at spatial infinity. Both EGD parameters ($\ell$ and $\chi$) modify the height and position of this barrier as their values increase from $0.4$ to $1.0$.

\begin{figure}[t]
    \centering
    \includegraphics[width=0.8\linewidth]{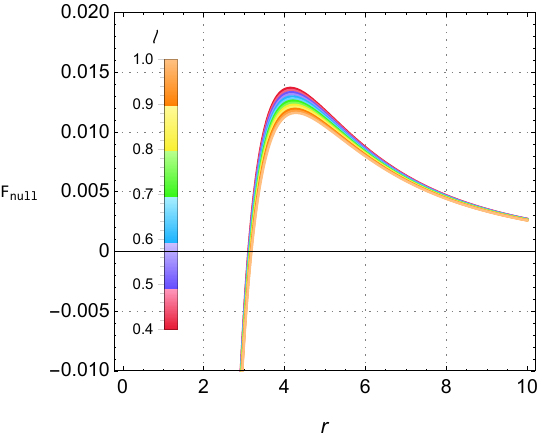}\qquad    \includegraphics[width=0.8\linewidth]{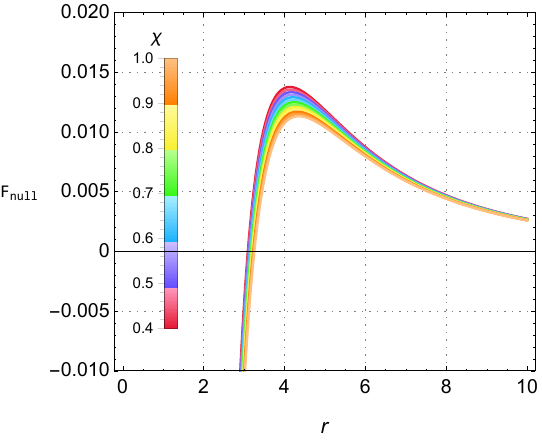}
    \caption{\footnotesize Behavior of the effective radial force 
$F^{\rm eff}_{\rm null}=-V'_{\rm null}/2$ as a function of the radial coordinate for varying EGD parameters $(\ell,\chi)$. The upper panel corresponds to fixed $\chi=0.5$, while the lower panel corresponds to fixed $\ell/M=0.5$, with $M=1$ and $L=2$. The dashed vertical lines indicate the outer event horizons, and only the exterior region $r\geq r_h$ is shown.}
    \label{fig:force-null}
\end{figure}

Now, we calculate the effective radial-force function for photons moving in the gravitational field produced by the chosen BH and demonstrate how the EGD influences this function compared to the standard one. Thus,
\begin{eqnarray}\nonumber
F^{\rm eff}_{\rm null}&=&-\frac{1}{2}\frac{dV_{null}}{dr}=\frac{L^2}{r^3}\bigg(1-\frac{3\ell \chi}{2r}-\frac{3M}{r}  +\frac{4M\ell \chi}{r^2} \\\label{force2}
&& - \left(1+\frac{3M}{r}\right)\frac{\chi}{2}\ee^{-r/M}\bigg).
\end{eqnarray}
In the limit $\chi=\ell=0$, the above equation reduces to the effective radial-force function for the standard Schwarzschild BH solution. In Fig. \ref{fig:force-null}, we present a plot illustrating the behavior of this function as a function of the radial coordinate, under variations of the EGD parameters $(\ell,\chi)$. The figure shows that these parameters introduce a non-trivial effective-force profile. The upper panel corresponds to fixed $\chi=0.5$, while the lower panel corresponds to fixed $\ell/M=0.5$. The sign of $F^{\rm eff}_{\rm null}$ describes the direction associated with the gradient of the effective potential and should not be interpreted directly as a physical repulsive gravitational force.

\subsection{Photon Sphere and Shadow Radius}

Circular photon orbits sit at the extrema of $V_{null}(r)$ where 
\begin{equation}
    \frac{d}{dr}\left( V_{null}(r)\right)=0 \quad \text{or} \quad rf'(r)-2f(r)=0.
\end{equation}
Explicitly, the photon sphere equation is
\begin{equation}
2r(3M-r)-8\chi\ell M+3\chi\ell r +\chi r(r+3M)e^{-r/M}=0.\label{eqphoton}
\end{equation}
The shadow radius as seen by a distant observer is given by the impact parameter corresponding to the unstable circular photon orbit
\begin{equation}\label{p28}
R_{\text{sh}} = b_c = \frac{r_{\text{ph}}}{\sqrt{f(r_{\text{ph}})}}.
\end{equation}
Thus,
\begin{equation}\label{p30}
R_{\text{sh}} = \frac{r_{\text{ph}}}{\sqrt{1-\frac{2M+\chi\ell}{r_{\text{ph}}}+\frac{2\chi\ell M}{r_{\text{ph}}^2}
-\frac{\chi
M}{r_{\text{ph}}}\ee^{-r_{ph}/M}}}.
\end{equation}

\begin{table}[htb!]
\centering
\resizebox{\columnwidth}{!}{%
\begin{tabular}{c|cc|cc|cc|cc}
\hline\hline
$\chi\rightarrow$
& \multicolumn{2}{c|}{$0.2$}
& \multicolumn{2}{c|}{$0.4$}
& \multicolumn{2}{c|}{$0.6$}
& \multicolumn{2}{c}{$0.8$}
\\ \hline
$\ell/M\downarrow$
& $\dfrac{r_{\rm ph}}{\mathcal{M}}$
& $\dfrac{R_{\rm sh}}{\mathcal{M}}$
& $\dfrac{r_{\rm ph}}{\mathcal{M}}$
& $\dfrac{R_{\rm sh}}{\mathcal{M}}$
& $\dfrac{r_{\rm ph}}{\mathcal{M}}$
& $\dfrac{R_{\rm sh}}{\mathcal{M}}$
& $\dfrac{r_{\rm ph}}{\mathcal{M}}$
& $\dfrac{R_{\rm sh}}{\mathcal{M}}$
\\ \hline
$0.4$
& 2.9266 & 5.0888
& 2.8595 & 4.9912
& 2.7980 & 4.9024
& 2.7416 & 4.8216
\\
$0.6$
& 2.8786 & 5.0271
& 2.7728 & 4.8811
& 2.6805 & 4.7548
& 2.5996 & 4.6454
\\
$0.8$
& 2.8326 & 4.9683
& 2.6932 & 4.7809
& 2.5768 & 4.6267
& 2.4795 & 4.4999
\\
$1.0$
& 2.7886 & 4.9122
& 2.6201 & 4.6899
& 2.4854 & 4.5158
& 2.3783 & 4.3804
\\ \hline\hline
\end{tabular}%
}
\caption{Dimensionless photon-sphere radius $r_{\rm ph}/\mathcal{M}$ 
and shadow radius $R_{\rm sh}/\mathcal{M}$ for different values of 
$\chi$ and $\ell/M$. Here, $M=1$ is the seed mass and 
$\mathcal{M}=M+\chi\ell/2$ is the physical asymptotic mass. All 
configurations satisfy $\ell/M\geq0.2888$.}
\label{tableShadow}
\end{table}
\begin{figure}[t]
    \centering    \includegraphics[width=0.8\linewidth]{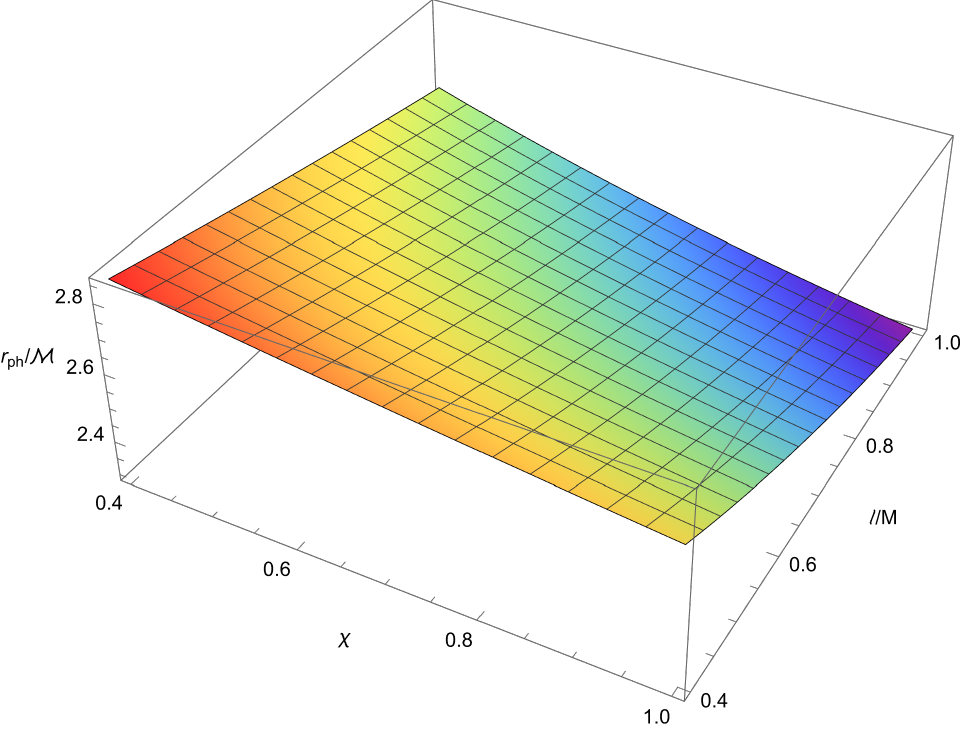}\qquad \includegraphics[width=0.8\linewidth]{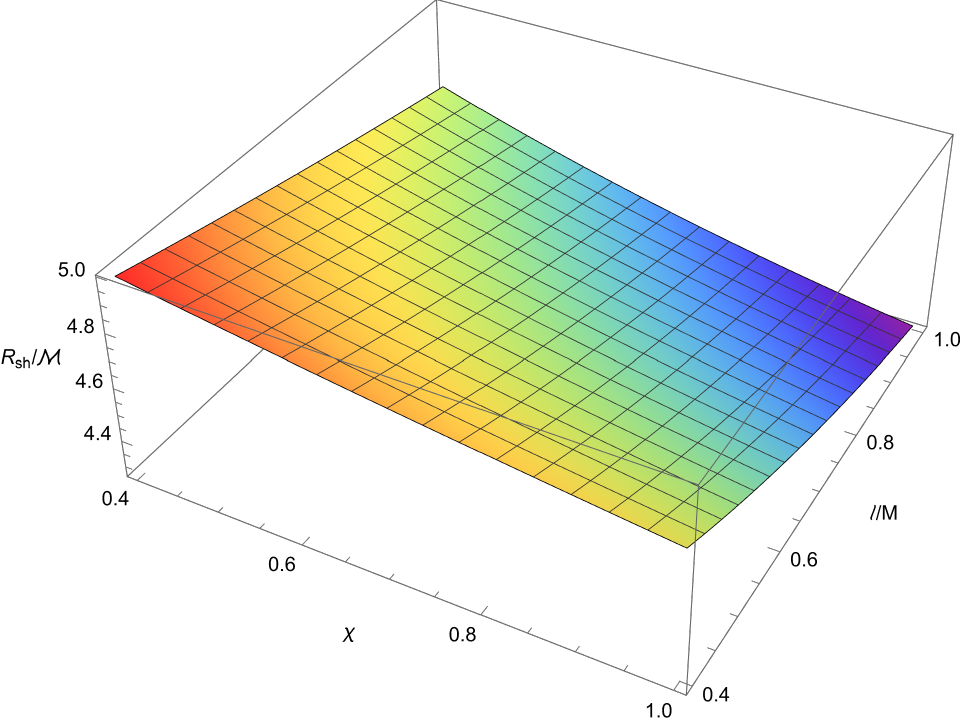}
    \caption{\footnotesize Three-dimensional behavior of the dimensionless photon-sphere radius $r_{\rm ph}/\mathcal{M}$ and shadow radius $R_{\rm sh}/\mathcal{M}$ for varying EGD parameters $(\ell/M,\chi)$, where $\mathcal{M}=M+\chi\ell/2$ is the asymptotic mass.}
    \label{fig:shadow2}
\end{figure}
Table \ref{tableShadow} and Figure \ref{fig:shadow2} illustrate the dependence of the normalized photon sphere radius ($r_{\text{ph}}/\mathcal{M}$) and the BH shadow radius ($R_{\text{sh}}/\mathcal{M}$) on the EGD model parameters $\chi$ and $\ell/M$ (for $M=1$). Due to the presence of exponential terms in Eq. (\ref{eqphoton}), the photon sphere radius cannot be solved analytically; hence, numerical methods were employed to determine both $r_{\text{ph}}$ and $R_{\text{sh}}$. As shown in Table \ref{tableShadow}, both normalized radii exhibit a monotonically decreasing trend with respect to both parameters. For instance, at fixed $\ell/M=0.4$, increasing $\chi$ from $0.2$ to $0.8$ decreases $r_{\text{ph}}/\mathcal{M}$ from $2.9266$ to $2.7416$ and $R_{\text{sh}}/\mathcal{M}$ from $5.0888$ to $4.8216$. Similarly, at fixed $\chi=0.8$, increasing $\ell/M$ from $0.4$ to $1.0$ decreases these quantities from $2.7416$ to $2.3783$ and from $4.8216$ to $4.3804$, respectively.
The 3D surface plots in Figure \ref{fig:shadow2} visually reinforce these numerical trends across the parameter space $\chi\in[0.2,0.8]$ and $\ell/M\in[0.4,1.0]$. Both surfaces display a smooth downward trend, showing that the normalized photon sphere and shadow radii decrease as the EGD parameters increase. In particular, both normalized radii remain below their corresponding Schwarzschild values in the parameter region considered.

\subsection{ Motion of massive particles}

In this section, we investigate the geodesic motion of massive particles in the hairy BH spacetime described by the metric function \eqref{METRIC}. The study of timelike geodesics is essential for understanding accretion disk physics, determining the ISCO, and characterizing the gravitational field in the strong-field regime. We derive the effective potential governing particle motion, analyze circular orbits, and compute the ISCO radius, which plays a crucial role in astrophysical observations. In the case of a massive test particle moving in the equatorial plane, the normalization condition $g_{\beta\alpha}\,u^\beta u^\alpha = 1$ gives
\begin{equation}\label{radmassive}
\dot{r}^2 = \mathrm E^2 - V_{\mathrm{eff}}(r)\,,
\end{equation}
where 
\begin{eqnarray}
V_{\rm eff}(r) &=& \left(1-\frac{2M+\chi\ell}{r} + \frac{2\chi\ell M}{r^{2}}
-\frac{\chi M}{r}e^{-r/M}\right)\nonumber\\
&& \times \left(1+\frac{L^{2}}{r^{2}}\right).\label{potential22}    
\end{eqnarray}

For a general static and spherically symmetric spacetime, the conserved angular momentum and energy corresponding to circular timelike orbits are obtained as
\begin{widetext}
\begin{align}
L^{2}
&=
\frac{r^{3}f'(r)}
{2f(r)-rf'(r)}=\frac{r^{2}\left[r(M+r)\chi + e^{r/M}\left(2Mr + \ell(r-4M)\chi\right)\right]}{e^{r/M}\left(2r^{2}-6Mr + 8\chi\ell M - 3\chi\ell r\right) - r(r+3M)\chi},
\label{Lsquare}
\\
E^{2}
&=
\frac{2f^{2}(r)}
{2f(r)-rf'(r)}=\frac{ 2e^{r/M}\left( Mr\chi+e^{r/M}(2M-r)(r-\chi \ell)  \right)^2}{r^2\left(e^{r/M}\left(2r^{2}-6Mr + 8\chi \ell M - 3\chi\ell r\right) - r(r+3M)\chi\right)}.
\label{Esquare}
\end{align}
\end{widetext}
For positive values of $E^2$ and $L^2$, the condition $2f(r)>rf'(r)$ must be satisfied, which sets a lower bound on the radii that can support circular motion.

\begin{figure}[t]
    \centering    \includegraphics[width=0.8\linewidth]{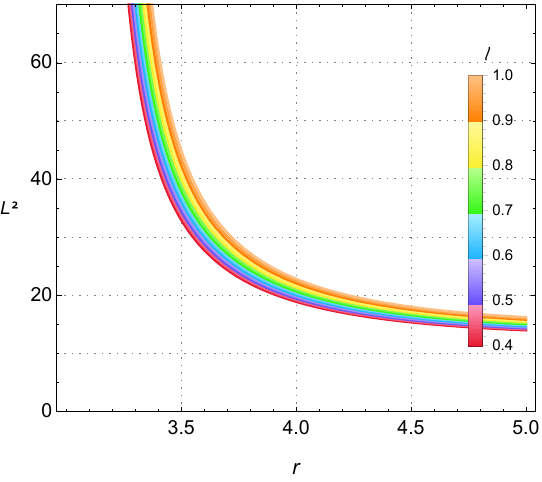}\qquad \includegraphics[width=0.8\linewidth]{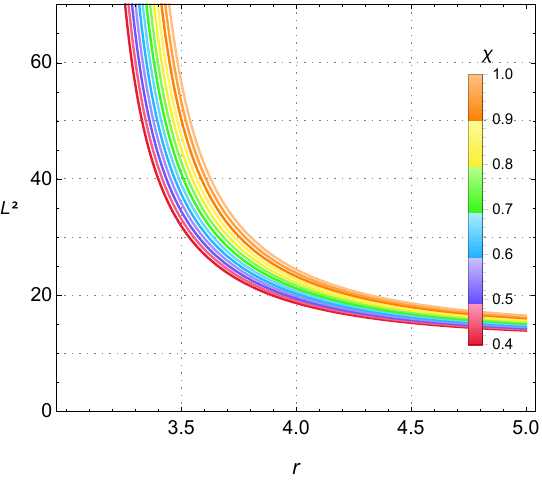}
    \caption{\footnotesize Behavior of the squared specific angular
momentum $L^{2}$ as a function of $r$ for varying EGD parameters
$(\ell/M,\chi)$, with $M=1$. The
upper panel corresponds to fixed $\chi=0.5$, while the lower
panel corresponds to fixed $\ell/M=0.5$.}
    \label{fig:angular2}
\end{figure}

\begin{figure}[t]
    \centering    \includegraphics[width=0.8\linewidth]{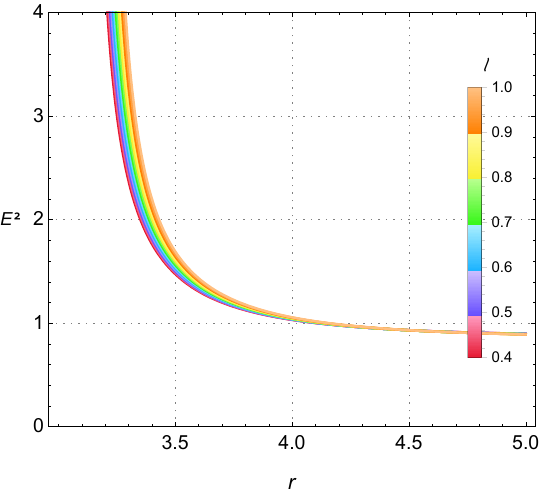}\qquad \includegraphics[width=0.8\linewidth]{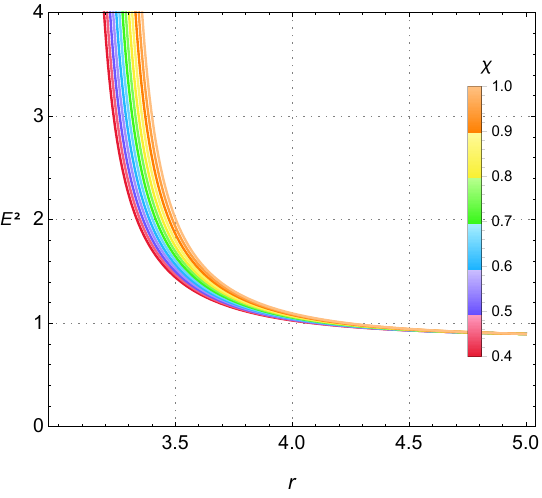}
    \caption{\footnotesize Behavior of the squared specific energy $E^{2}$ as a
function of $r$ for varying EGD parameters $(\ell/M,\chi)$, with
$M=1$. The upper panel corresponds to fixed $\chi=0.5$, while the
lower panel corresponds to fixed $\ell/M=0.5$.}
    \label{fig:energy2}
\end{figure}
Figure \ref{fig:angular2} illustrates the behavior of the squared specific angular momentum $L^2$ as a function of $r$ for $M=1$. The upper panel displays curves for varying values of $\ell/M$ from $0.4$ to $1.0$ at fixed $\chi=0.5$, while the lower panel shows curves for varying values of $\chi$ from $0.4$ to $1.0$ at fixed $\ell/M=0.5$. In the radial interval shown, $L^2$ decreases as $r$ increases. A similar trend is observed in Figure \ref{fig:energy2} for $E^2$ within the displayed interval. These plots illustrate only the radial behavior over the interval shown; the ISCO values are obtained independently from the marginal stability condition and are reported in Table \ref{isco3}.

The ISCO is determined from the marginal stability condition $\mathrm{d}^2 V_{\mathrm{eff}}/\mathrm{d}r^2 = 0$. The radiative efficiency of an accretion disk is
\begin{equation}\label{m43}
\epsilon = 1 - E_{\text{ISCO}}.
\end{equation} 
Equation (\ref{m43}) measures how efficiently the gravitational energy of infalling matter is converted into radiation before the matter crosses the ISCO. For Schwarzschild, $\epsilon \approx 5.72\%$. For our hairy BH metric obtained through the EGD procedure,
\begin{equation}\label{m44}
\epsilon = 1 - \sqrt{\frac{2f(r_{\text{ISCO}})^2}{2f(r_{\text{ISCO}}) - r_{\text{ISCO}} f'(r_{\text{ISCO}})}}.
\end{equation}

The stability condition can be expressed in terms of the metric function as
\begin{equation}\label{m38}
3\, f\, f'- 2\, r\, (f')^2+r\, f\, f''= 0.
\end{equation}
From Eq. (\ref{m38}), it is clear that the ISCO radius $r_{ISCO}$ depends on the EGD parameters
$(\ell,\chi)$. The
ISCO radius reduces to the Schwarzschild
result $r_{ISCO}=6M$ when $\ell= \chi=0$.
\begin{table*}[t]
\centering
\resizebox{\textwidth}{!}{%
\begin{tabular}{c|ccc|ccc|ccc|ccc}
\hline\hline
$\ell/M\rightarrow$
& \multicolumn{3}{c|}{$0.4$}
& \multicolumn{3}{c|}{$0.6$}
& \multicolumn{3}{c|}{$0.8$}
& \multicolumn{3}{c}{$1.0$}
\\ \hline
$\chi\downarrow$
& $\dfrac{r_{\rm ISCO}}{\mathcal{M}}$
& $E_{\rm ISCO}$
& $\epsilon(\%)$
& $\dfrac{r_{\rm ISCO}}{\mathcal{M}}$
& $E_{\rm ISCO}$
& $\epsilon(\%)$
& $\dfrac{r_{\rm ISCO}}{\mathcal{M}}$
& $E_{\rm ISCO}$
& $\epsilon(\%)$
& $\dfrac{r_{\rm ISCO}}{\mathcal{M}}$
& $E_{\rm ISCO}$
& $\epsilon(\%)$
\\ \hline
$0.2$
& 5.8123 & 0.9410 & 5.90\%
& 5.7059 & 0.9400 & 6.00\%
& 5.6047 & 0.9390 & 6.10\%
& 5.5085 & 0.9380 & 6.20\%
\\
$0.4$
& 5.6416 & 0.9392 & 6.08\%
& 5.4522 & 0.9373 & 6.27\%
& 5.2808 & 0.9355 & 6.45\%
& 5.1256 & 0.9338 & 6.62\%
\\
$0.6$
& 5.4860 & 0.9375 & 6.25\%
& 5.2327 & 0.9349 & 6.51\%
& 5.0144 & 0.9324 & 6.76\%
& 4.8265 & 0.9301 & 6.99\%
\\
$0.8$
& 5.3440 & 0.9359 & 6.41\%
& 5.0424 & 0.9326 & 6.74\%
& 4.7952 & 0.9295 & 7.05\%
& 4.5937 & 0.9269 & 7.31\%
\\
\hline\hline
\end{tabular}%
}
\caption{Dimensionless ISCO radius $r_{\rm ISCO}/\mathcal{M}$,
specific energy $E_{\rm ISCO}$, and radiative efficiency
$\epsilon$ for different values of $\chi$ and $\ell/M$. Here,
$M=1$ is the seed mass and $\mathcal{M}=M+\chi\ell/2$ is the
physical asymptotic mass. All configurations satisfy
$\ell/M\geq0.2888$.}
\label{isco3}
\end{table*}
To systematically quantify the influence of the EGD parameters on orbital dynamics and energetic efficiency around the hairy BH, numerical values of the normalized ISCO radius $r_{ISCO}/\mathcal{M}$, the corresponding specific energy $E_{ISCO}$, and the radiative efficiency $\epsilon(\%)$ are listed in Table \ref{isco3} for representative values of $\ell/M$ and $\chi$ (setting $M=1$). As expected, in the vanishing-hair limit $\chi\to0$, the parameters reduce to the standard Schwarzschild baseline ($r_{ISCO}/\mathcal{M}=6$, $E_{ISCO}\approx0.9428$, and $\epsilon\approx5.72\%$). Increasing either $\ell/M$ or $\chi$ leads to a monotonic decrease in $r_{ISCO}/\mathcal{M}$, shifting the innermost stable orbit inward in units of the asymptotic mass. Concurrently, the specific energy $E_{ISCO}$ decreases, while the accretion efficiency $\epsilon$ increases. For example, for $\chi=0.8$, increasing $\ell/M$ from $0.4$ to $1.0$ decreases $r_{ISCO}/\mathcal{M}$ from $5.3440$ to $4.5937$ and increases the efficiency from $6.41\%$ to $7.31\%$. This enhanced efficiency allows infalling matter to release a larger fraction of its rest-mass energy prior to crossing the ISCO.

\begin{figure}[t]
    \centering    \includegraphics[width=0.8\linewidth]{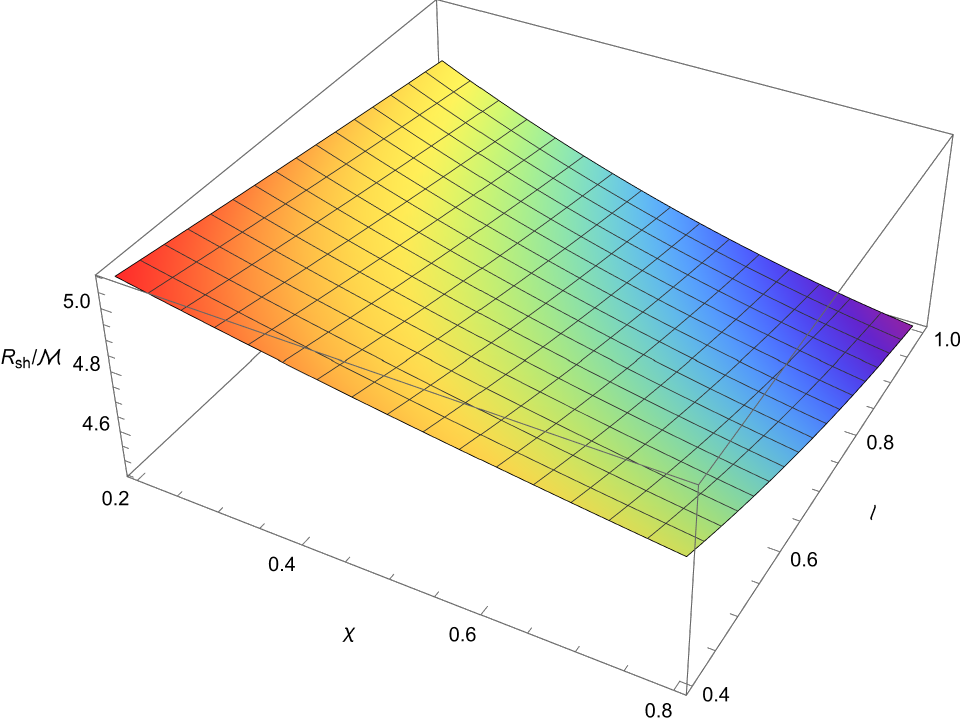}\qquad \includegraphics[width=0.8\linewidth]{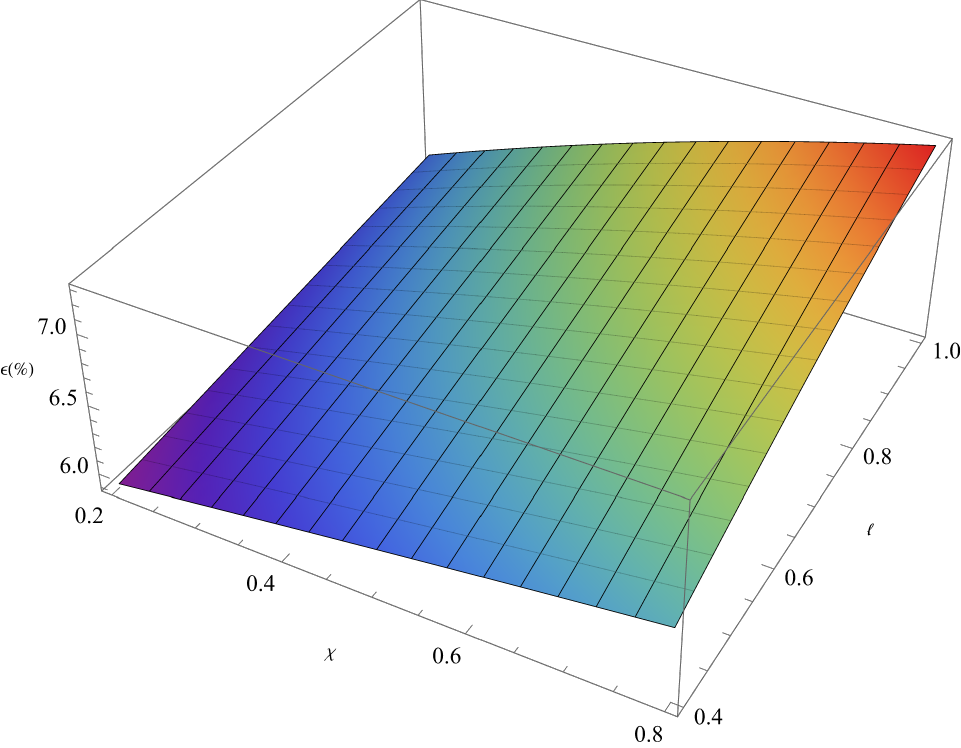}
    \caption{\footnotesize Three-dimensional behavior of the normalized ISCO radius $r_{\rm ISCO}/\mathcal{M}$ and radiative efficiency
$\epsilon(\%)$ for varying EGD parameters $(\ell/M,\chi)$.}
    \label{fig:ISCO2}
\end{figure}
The continuous global trends and interplay between the decoupling parameters $(\ell/M,\chi)$ are illustrated in the three-dimensional surface plots in Figure \ref{fig:ISCO2}. The left panel demonstrates that the normalized ISCO radius $r_{ISCO}/\mathcal{M}$ decreases monotonically with increasing values of both parameters. Conversely, the right panel shows that the accretion efficiency $\epsilon(\%)$ increases towards the upper parameter boundaries. Together, these surfaces show that the gravitational hair shifts the normalized ISCO inward while increasing the energy output of the accretion disk.

\subsection{Particle Trajectories}

In this subsection, we analyze the orbital shape of test particles moving around the hairy BH obtained by EGD and examine how the EGD parameters $(\ell,\chi)$ modify
the trajectory paths.
We start with the radial equation of motion (\ref{radmassive}) 
rewritten in terms of the azimuthal angle $\phi$: \begin{equation}
    \left( \frac{1}{r^2}\frac{dr}{d\phi} \right)^2 = \frac{{E}^2}{{L}^2} - \frac{f(r)}{{L}^2} - \frac{f(r)}{r^2}.
\end{equation}
Using the reciprocal substitution $u = 1/r$, we obtain 
\begin{eqnarray}\nonumber
    && \left( \frac{du}{d\phi} \right)^2 + u^2 = \frac{{E}^2 - 1}{{L}^2} + \frac{2M + \chi\ell}{{L}^2} u - \frac{2\chi\ell M}{{L}^2} \\\nonumber&& u^2 + (2M + \chi\ell) u^3 - 2\chi\ell M u^4 + \chi M \left(\frac{u}{{L}^2} + u^3\right) e^{-\frac{1}{M u}}.
\end{eqnarray}
To eliminate the squared derivative $\left(\frac{du}{d\phi}\right)^2$, we differentiate the first-order equation with respect to $\phi$ and obtain the modified second-order nonlinear ordinary differential equation (ODE)
\begin{widetext} 
\begin{equation}
    \frac{d^2 u}{d\phi^2} + u = \frac{2M + \chi\ell}{2{L}^2} - \frac{2\chi\ell M}{{L}^2} u + \frac{3}{2}(2M + \chi\ell) u^2 - 4\chi\ell M u^3 + \frac{\chi M}{2} \left[ \frac{1}{{L}^2} + 3u^2 + \frac{1}{M}\left( \frac{1}{{L}^2 u} + u \right) \right] e^{-\frac{1}{M u}}.\label{trajectory2}
\end{equation}
\end{widetext}
To analyze the particle trajectory, we solve Eq. (\ref{trajectory2}) numerically with initial conditions $u(0)=0.15$, $u'(0)=0$, setting $M=1$.

\begin{figure*}[t]
    \centering  \includegraphics[width=\textwidth]{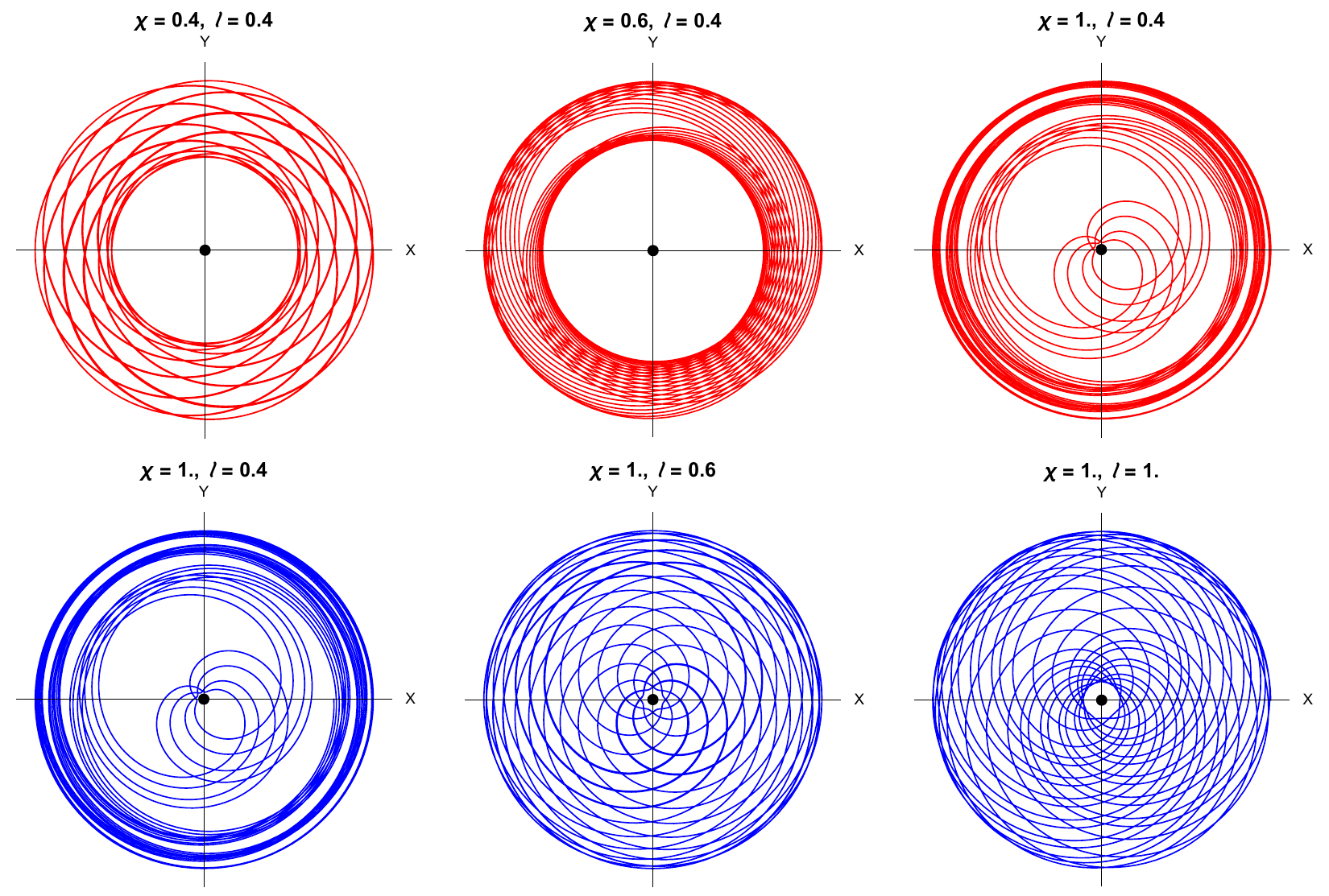}
    \caption{\footnotesize Particle trajectory paths for varying EGD parameters $\chi$ and $\ell/M$, with the boundary conditions $u(0)=0.15$, $u'(0)=0$ and $M=1,L=1$.}
    \label{fig:trajectory}
\end{figure*}

Figure \ref{fig:trajectory} represents the particle trajectories in the $X-Y$ orbital plane and illustrates the influence of the EGD parameters $\chi$ and $\ell$ on test-particle motion around the hairy BH. The central black dot indicates the position of the BH at the origin $(0,0)$. The top row (red curves) displays the evolution of the orbits as $\chi$ increases from $0.4$ to $1.0$ at fixed $\ell/M=0.4$. The bottom row (blue curves) demonstrates the effect of increasing $\ell/M$ from $0.4$ to $1.0$ at fixed $\chi=1$. The changes in the rosette patterns show qualitatively that both parameters modify the orbital structure. However, a quantitative statement about perihelion precession requires a calculation from successive pericenters and is not inferred directly from these plots.

\section{Thermodynamic Properties}\label{section V}

In this section, we investigate the thermodynamic properties of the 
charged-like DEC branch introduced in the previous sections. Before defining the thermal response, it is necessary to specify the thermodynamic family along which the black-hole parameters are varied. Throughout this section, we impose the charged-like DEC branch constraint
\[
Q^{2}=2\chi\ell M,
\]
and choose the effective charge $Q$ and the hair scale $\ell$ as the 
quantities held fixed. Consequently, the deformation parameter $\chi$ is 
not an independent thermodynamic variable but satisfies
\[
\chi=\frac{Q^{2}}{2\ell M}.
\]
Thus, when the event-horizon radius $r_h$ is varied, the mass parameter 
$M=M(r_h;Q,\ell)$ changes along the physical horizon family, whereas 
$Q$ and $\ell$ remain fixed.

Using the constraint $Q^{2}=2\chi\ell M$, the metric function of the 
charged-like DEC branch can be rewritten as
\begin{equation}
f(r)=
1-\frac{2M+\dfrac{Q^{2}}{2M}}{r}
+\frac{Q^{2}}{r^{2}}
-\frac{Q^{2}}{2\ell r}\exp\left(-\frac{r}{M}\right).
\label{eq:thermo_metric}
\end{equation}

The event horizon is determined by $f(r_h)=0$. We therefore introduce
\begin{eqnarray}
\mathcal{H}(r_h,M;Q,\ell) &\equiv& 1-\frac{2M+\dfrac{Q^{2}}{2M}}{r_h} +\frac{Q^{2}}{r_h^{2}}\nonumber\\
&& -\frac{Q^{2}}{2\ell r_h} \exp\left(-\frac{r_h}{M}\right) = 0.\label{eq:horizon_family}
\end{eqnarray}
For fixed $(Q,\ell)$, Eq.~\eqref{eq:horizon_family} implicitly determines
the mass parameter as $M=M(r_h;Q,\ell)$. Only solutions corresponding to 
the outer event horizon and satisfying the physically admissible DEC 
domain are retained in the following analysis.

The Hawking temperature is determined by the surface gravity,
\begin{equation}
T_H=
\frac{1}{4\pi}
\left.
\frac{\partial f(r)}{\partial r}
\right|_{r=r_h},
\label{eq:hawking_definition}
\end{equation}
which gives
\begin{align}
T_H=
\frac{1}{4\pi}
\Bigg[
&\frac{2M+\dfrac{Q^{2}}{2M}}{r_h^{2}}
-\frac{2Q^{2}}{r_h^{3}}
\nonumber\\
&+
\frac{Q^{2}}{2\ell}
\exp\left(-\frac{r_h}{M}\right)
\left(
\frac{1}{r_h^{2}}
+\frac{1}{Mr_h}
\right)
\Bigg].
\label{eq:hawking_Qell}
\end{align}
Here, the partial derivative in Eq.~\eqref{eq:hawking_definition} is 
evaluated for a single geometry with $(M,Q,\ell)$ fixed. However, when 
different black-hole solutions along the horizon family are compared, 
$M$ changes according to Eq.~\eqref{eq:horizon_family}.

It is also useful to reconsider the limit $r_h=2M$. Substitution of 
$r_h=2M$ into Eq.~\eqref{eq:horizon_family} yields
\[
-\frac{Q^{2}}{4\ell M}\exp(-2)=0,
\]
which, for finite $M$ and $\ell$, requires
\[
Q=0.
\]
The charged-like DEC branch constraint then gives $\chi=0$. Therefore, $r_h=2M$
corresponds only to the Schwarzschild limit and does not represent a 
non-trivial hairy configuration. In this limit,
\begin{equation}
T_H\longrightarrow
\frac{1}{8\pi M}
=
T_{\rm Schw}.
\label{eq:schwarzschild_temperature}
\end{equation}

\begin{figure}[t]
   \centering  \includegraphics[width=0.950\linewidth]{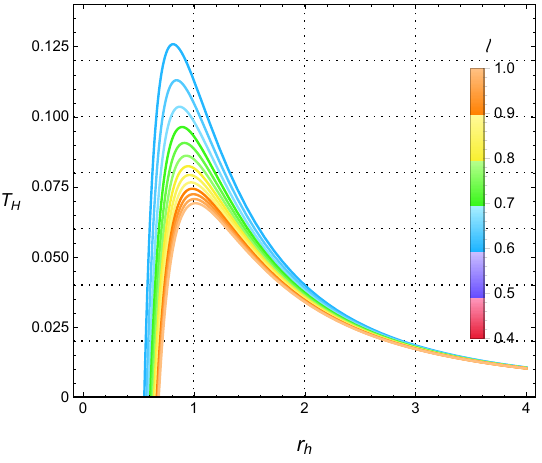}
  \caption{\footnotesize Hawking temperature $T_H$ as a function of the outer event-horizon radius $r_h$ for the charged-like hairy black hole in the fixed-$(Q,\ell)$ thermodynamic ensemble with $Q=1$. For each value of $r_h$, the seed mass $M=M(r_h;Q,\ell)$ is obtained from the horizon equation $f(r_h)=0$. Only the outer-horizon branch is considered in the thermodynamic analysis.}
   \label{fig:temp1}
\end{figure}

Figure \ref{fig:temp1} illustrates the behavior of the Hawking temperature $T_H$ as a function of the horizon radius $r_h$ for fixed effective charge $Q$. The curves display a family of thermodynamic profiles color-coded by the hair parameter $\ell$, showing how varying $\ell$ modifies the temperature profile of the black hole.

The entropy obeys the Bekenstein--Hawking area law,
\begin{equation}
S=\pi r_h^{2}.
\label{eq:entropy}
\end{equation}

Before defining the heat capacity, we establish the differential 
thermodynamic structure of the selected solution family. From the 
asymptotic $1/r$ behavior of the metric, the physical asymptotic mass is
\begin{equation}
\mathcal{M}
=
M+\frac{\chi\ell}{2}
=
M+\frac{Q^{2}}{4M}.
\label{eq:asymptotic_mass}
\end{equation}
The distinction between $\mathcal{M}$ and the seed mass parameter $M$ is
important because $M$ also appears explicitly in the exponential 
deformation.

Differentiating the horizon condition
$\mathcal{H}(r_h,M;Q,\ell)=0$ gives
\begin{equation}
\mathcal{H}_{r_h}\,dr_h+
\mathcal{H}_M\,dM+
\mathcal{H}_Q\,dQ+
\mathcal{H}_{\ell}\,d\ell=0,
\label{eq:horizon_differential}
\end{equation}
where $\mathcal{H}_x\equiv\partial \mathcal{H}/\partial x$. In the fixed-$(Q,\ell)$ ensemble,
$dQ=d\ell=0$, and therefore
\begin{equation}
\left(
\frac{dM}{dr_h}
\right)_{Q,\ell}
=
-\frac{\mathcal{H}_{r_h}}{\mathcal{H}_M}.
\label{eq:dMdrh}
\end{equation}
For completeness,
\begin{equation}
\mathcal{H}_M=
-\frac{2}{r_h}
+\frac{Q^{2}}{2M^{2}r_h}
-\frac{Q^{2}}{2M^{2}\ell}
\exp\left(-\frac{r_h}{M}\right),
\label{eq:FM}
\end{equation}
whereas $\mathcal{H}_{r_h}=4\pi T_H$.

Because the hairy deformation depends explicitly on the seed mass $M$, 
the variation of the asymptotic mass does not reduce identically to the 
standard Schwarzschild relation $d\mathcal{M}=T_HdS$. Instead, the 
horizon variation can be written in the modified first-law form
\begin{equation}
\Gamma\,d\mathcal{M}
=
T_H\,dS
+\Phi_Q\,dQ
+\Psi_{\ell}\,d\ell ,
\label{eq:modified_first_law}
\end{equation}
where
\begin{equation}
\Gamma
=
-\frac{r_h\mathcal{H}_M}
{2\left(1-\dfrac{Q^{2}}{4M^{2}}\right)}
=
1+
\frac{Q^{2}r_h
\exp\left(-r_h/M\right)}
{\ell\left(4M^{2}-Q^{2}\right)}.
\label{eq:Gamma}
\end{equation}
The quantities $\Phi_Q$ and $\Psi_{\ell}$ denote the thermodynamic 
potentials conjugate to the effective charge and the hair scale, 
respectively. They may be expressed as
\begin{equation}
\Phi_Q=
\Gamma
\left[
\frac{Q}{2M}
-
\left(
1-\frac{Q^{2}}{4M^{2}}
\right)
\frac{\mathcal{H}_Q}{\mathcal{H}_M}
\right],
\label{eq:PhiQ}
\end{equation}
and
\begin{equation}
\Psi_{\ell}
=
\frac{Q^{2}}{4\ell^{2}}
\exp\left(-\frac{r_h}{M}\right).
\label{eq:Psiell}
\end{equation}
In the Schwarzschild limit $Q\rightarrow0$, one has
$\Gamma\rightarrow1$, and the standard Schwarzschild thermodynamic 
relation is recovered.

Having specified the ensemble and the corresponding horizon variation,
we now define the heat capacity at fixed effective charge $Q$ and fixed
hair scale $\ell$. From Eq.~\eqref{eq:modified_first_law}, the appropriate
energy-response heat capacity is
\begin{equation}
C_{Q,\ell}
\equiv
\left(
\frac{d\mathcal{M}}{dT_H}
\right)_{Q,\ell}
=
\frac{T_H}{\Gamma}
\left(
\frac{dS}{dT_H}
\right)_{Q,\ell}.
\label{eq:heat_capacity_definition}
\end{equation}
Using $S=\pi r_h^{2}$, this becomes
\begin{equation}
C_{Q,\ell}
=
\frac{2\pi r_hT_H}
{\Gamma
\left(\dfrac{dT_H}{dr_h}\right)_{Q,\ell}}.
\label{eq:heat_capacity}
\end{equation}
The temperature derivative appearing in Eq.~\eqref{eq:heat_capacity}
is a total derivative along the physical fixed-$(Q,\ell)$ horizon family,
\begin{equation}
\left(
\frac{dT_H}{dr_h}
\right)_{Q,\ell}
=
\left(
\frac{\partial T_H}{\partial r_h}
\right)_{M,Q,\ell}
+
\left(
\frac{\partial T_H}{\partial M}
\right)_{r_h,Q,\ell}
\left(
\frac{dM}{dr_h}
\right)_{Q,\ell},
\label{eq:total_temperature_derivative}
\end{equation}
with
\[
\left(
\frac{dM}{dr_h}
\right)_{Q,\ell}
=
-\frac{\mathcal{H}_{r_h}}{\mathcal{H}_M}.
\]
Equation~\eqref{eq:heat_capacity}, together with
Eqs.~\eqref{eq:horizon_family} and
\eqref{eq:total_temperature_derivative}, is used for the numerical
heat-capacity analysis. This construction differs from varying $r_h$ at
fixed $(M,\ell)$, because both $M$ and $\chi$ now evolve consistently
along the same physical black-hole family while $Q$ and $\ell$ remain
fixed. Explicitly, the heat capacity is \begin{multline}
C_{Q,\ell} = 2\pi r_h \left(1 - \frac{Q^2}{4M^2}\right) 
\left[ 1 + \frac{Q^{2}r_h e^{-r_h/M}}{\ell\left(4M^{2}-Q^{2}\right)} \right]^{-1} \\
\times \frac{ \left(2M + \dfrac{Q^2}{2M}\right)r_h - 2Q^2 + \dfrac{Q^2 r_h^2}{2\ell}e^{-\frac{r_h}{M}}\left(\dfrac{1}{r_h^2}+\dfrac{1}{Mr_h}\right) }
{ -\dfrac{4M + \frac{Q^2}{M}}{r_h} + \dfrac{6Q^2}{r_h^2} - \dfrac{Q^2}{2\ell M} e^{-\frac{r_h}{M}} \left( 1 + \dfrac{3r_h}{M} + \dfrac{M}{r_h} \right) }.
\label{eq:heat_capacity_final}
\end{multline}

\begin{figure}[t]
   \centering    \includegraphics[width=0.95\linewidth]{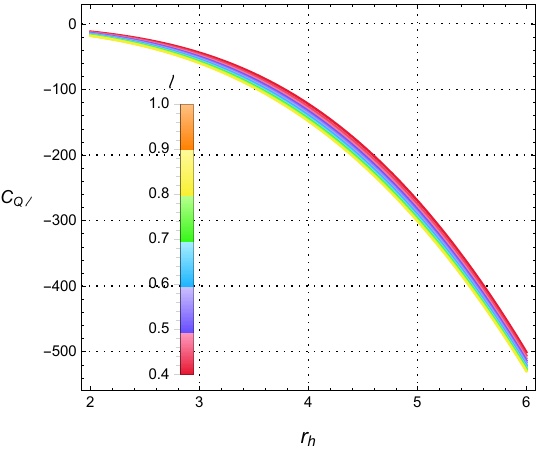}
   \caption{\footnotesize Heat capacity $C_{Q,\ell}$ as a function of the outer event-horizon radius $r_h$ for the charged-like hairy black hole in the fixed-$(Q,\ell)$ thermodynamic ensemble with $Q=1$. For each value of $r_h$, the seed mass $M=M(r_h;Q,\ell)$ is determined from the horizon equation $f(r_h)=0$. Only the outer-horizon branch is considered.}
   \label{fig:capacity}
\end{figure}

Figure~\ref{fig:capacity} shows the corrected heat capacity $C_{Q,\ell}$ given by Eq. (\ref{eq:heat_capacity_final}) over the thermodynamic domain considered. Since the charged-like DEC branch requires $r_h\geq2M$, only this physically admissible region is included.

\section{Conclusions}\label{section VI}

In this manuscript, we have studied the dynamics of the trajectories of massless and massive particles in the vicinity of a hairy black hole described by a geometric solution obtained by applying gravitational decoupling through extended geometric deformation to the Schwarzschild vacuum solution. This geometry satisfies the dominant energy condition (DEC) and the condition $Q^{2}=2\chi\ell M$. This choice is useful because it links the effective ``charge'' $Q$ with the gravitational hair parameters $\chi$ and $\ell$. This avoids treating $Q$ as an independent parameter and reduces the number of parameters in our analysis. Furthermore, it guarantees that \(Q\rightarrow0\) when
\(\chi\rightarrow0\) or \(\ell\rightarrow0\), although the latter is 
only a formal limit for the DEC branch considered here. This condition factorizes the Reissner--Nordstr\"om-type sector as
\begin{eqnarray}
1-\frac{2M+\chi\ell}{r} +\frac{2\chi\ell M}{r^{2}} = \left(1-\frac{2M}{r}\right)\left(1-\frac{\chi\ell}{r}\right),
\end{eqnarray}
which allows us to interpret the Reissner--Nordstr\"om-type sector as the combination of two scales, $2M$ and $\chi\ell$. In particular, under this condition, the effective charge never exceeds the asymptotic mass in the Reissner--Nordstr\"om-type sector. Indeed, using $\mathcal{M}=M+\frac{\chi\ell}{2}$, we obtain
\begin{eqnarray}
\mathcal{M}^{2}-Q^{2} = \left(M-\frac{\chi\ell}{2}\right)^{2} \geq 0,    
\end{eqnarray}
and hence $Q^{2}\leq \mathcal{M}^{2}$. Therefore, the Reissner--Nordstr\"om-type sector is automatically non-overcharged, it can be subextremal or extremal, but not superextremal.

It is important to mention that \(r=2M\) and \(r=\chi\ell\) are only the roots of the factorized sector and are not exact horizons of the complete metric, because the exponential term $-\frac{\chi M}{r}e^{-r/M}$ in the metric given by Eq. (\ref{METRIC}) shifts their positions.

For $\chi>0$, the condition $Q^{2}=2\chi\ell M$ simplifies the DEC to the constraint $\frac{\ell}{M} \geq 0.2888$. This constraint was used to determine the physically consistent parameter configurations for our physical system. The analysis of the metric function also shows the presence of inner and outer horizons for the physical configurations considered. Furthermore, the Schwarzschild horizon $r_h=2M$ is recovered only at the hairless boundary $\chi=0$, while a nonzero value of $\chi$ shifts the outer horizon away from the Schwarzschild radius.

Regarding the analysis of possible trajectories around the hairy black hole, for the case of massless particles we found an effective potential and the corresponding effective radial force function, which allowed us to obtain the condition $r f'(r)-2f(r)=0$ for an unstable circular orbit. This condition determines the radius of the photon sphere and the associated shadow radius. These quantities proved to be sensitive to the parameters $\chi$ and $\ell$, thus showing that the gravitational hair of the black hole modifies the propagation of photons in the strong-field region. It is important to mention that in our analysis we have distinguished the seed mass $M$ from the asymptotic mass $\mathcal{M}=M+\chi\ell/2$. In fact, when $M$ is fixed and $\chi$ and $\ell$ vary, $\mathcal{M}$ also changes; therefore, the variations in the characteristic radii cannot be attributed solely to the gravitational hair. Thus, to isolate this effect, the corresponding radii were normalized by the asymptotic mass. The results show that $r_{\rm ph}/\mathcal{M}$ and $R_{\rm sh}/\mathcal{M}$ decrease as either $\chi$ or $\ell/M$ increases in the parameter region considered.

In the case of massive particles, the specific energy and angular momentum associated with circular orbits were found, and the ISCO was determined from the marginal stability condition. We found that the presence of gravitational hair modifies the effective potential, the location of the ISCO, and the energy of the particles in the innermost stable orbit. In particular, $r_{\rm ISCO}/\mathcal{M}$ and $E_{\rm ISCO}$ decrease as the hair parameters increase, while the radiative efficiency increases. The effect of this gravitational hair is also visible in the numerical trajectories, which exhibit behavior different from the orbital structure of the Schwarzschild black hole. These results are interpreted using the asymptotic mass and within the parameter region allowed by the DEC.

Regarding the thermodynamic analysis of our physical system, we have analyzed the Hawking temperature and the thermal response of the geometry of our hairy black hole solution. For this purpose, we adopted the fixed-\((Q,\ell)\) ensemble, in which the event horizon equation determines the implicit variation of $M$ with $r_h$, while the condition $Q^{2}=2\chi\ell M$ determines the corresponding variation of \(\chi\). With this prescription, the heat capacity was evaluated along this consistent family of black hole solutions. In this framework, we obtained the heat capacity given by Eq. (\ref{eq:heat_capacity_definition}), which consistently accounts for the implicit dependence $M=M(r_h,Q,\ell)$. Therefore, all thermodynamic derivatives were evaluated with $Q$ and $\ell$ fixed, following the outer horizon branch that satisfies the DEC. For the configurations considered in Fig.~\ref{fig:capacity}, $C_{Q,\ell}$ remains negative throughout the displayed physical branch. Thus, these configurations are locally thermodynamically unstable, and no divergence or change of sign is observed within the parameter domain analyzed. In the limit $Q\rightarrow0$, and therefore $\chi\rightarrow0$ for finite and nonzero $\ell$, the Schwarzschild relations are recovered, including $C_{\rm Schw}=-8\pi M^{2}<0$, which corresponds to the well-known local thermodynamic instability of the Schwarzschild black hole.

In summary, it has been shown that the condition $Q^{2}=2\chi\ell M$ provides a simple framework in which the effective charge, gravitational hair, and asymptotic mass are directly related. The resulting geometry exhibits nontrivial optical, orbital, and thermodynamic properties and is consistent with the Schwarzschild black hole geometry in the vanishing-hair limit. This work can be extended to rotating geometries, gravitational lensing calculations, quasinormal modes, and observational constraints on the parameters associated with the gravitational hair.

\section*{Declaration of competing interest}

The authors declare no competing interests.

\section*{Conflict of Interest}

The authors declare that they have no known competing financial
interests or personal relationships that could have appeared to
influence the work reported in this paper

\section*{Data Availability Statement}

No datasets were generated, analyzed or used in this study.

\end{document}